\PassOptionsToPackage{unicode}{hyperref}
\PassOptionsToPackage{hyphens}{url}
\documentclass[
]{article}
\usepackage{amsmath,amssymb}
\usepackage{iftex}
\ifPDFTeX
  \usepackage[T1]{fontenc}
  \usepackage[utf8]{inputenc}
  \usepackage{textcomp} 
\else 
  \usepackage{unicode-math} 
  \defaultfontfeatures{Scale=MatchLowercase}
  \defaultfontfeatures[\rmfamily]{Ligatures=TeX,Scale=1}
\fi
\usepackage{lmodern}
\ifPDFTeX\else
\fi
\IfFileExists{upquote.sty}{\usepackage{upquote}}{}
\IfFileExists{microtype.sty}{
  \usepackage[]{microtype}
  \UseMicrotypeSet[protrusion]{basicmath} 
}{}
\makeatletter
\@ifundefined{KOMAClassName}{
  \IfFileExists{parskip.sty}{%
    \usepackage{parskip}
  }{
    \setlength{\parindent}{0pt}
    \setlength{\parskip}{6pt plus 2pt minus 1pt}}
}{
  \KOMAoptions{parskip=half}}
\makeatother
\usepackage{xcolor}
\usepackage{graphicx}
\makeatletter
\def\maxwidth{\ifdim\Gin@nat@width>\linewidth\linewidth\else\Gin@nat@width\fi}
\def\maxheight{\ifdim\Gin@nat@height>\textheight\textheight\else\Gin@nat@height\fi}
\makeatother
\setkeys{Gin}{width=\maxwidth,height=\maxheight,keepaspectratio}
\makeatletter
\def\fps@figure{htbp}
\makeatother
\ifLuaTeX
  \usepackage{selnolig}  
\fi
\IfFileExists{bookmark.sty}{\usepackage{bookmark}}{\usepackage{hyperref}}
\IfFileExists{xurl.sty}{\usepackage{xurl}}{} 
\hypersetup{
  hidelinks,
  pdfcreator={LaTeX via pandoc}}

\author{}
\date{}

\begin{document}

\hypertarget{team-composition-in-software-engineering-education}{%
\section*{Team Composition in Software Engineering
Education}\label{team-composition-in-software-engineering-education}}
\addcontentsline{toc}{section}{Team Composition in Software Engineering
Education}

Sajid Ibrahim Hashmi

M3S Research Unit, Faculty of Information Technology and Electrical
Engineering, University of Oulu, Finland, sajid.hashmi@oulu.fi

Jouni Markkula

M3S Research Unit, Faculty of Information Technology and Electrical
Engineering, University of Oulu, Finland, jouni.markkula@oulu.fi

One of the objectives of software engineering education is to make
students to learn essential teamwork skills. This is done by having the
students work in groups for course assignments. Student team composition
plays a vital role in this, as it significantly affects learning
outcomes, what is learned, and how. The study presented in this paper
aims to better understand the student team composition in software
engineering education and investigate the factors affecting it in the
international software engineering education context. Those factors
should be taken into consideration by software engineering teachers when
they design group work assignments in their courses. In this paper, the
initial findings of the ongoing Action research study are presented. The
results give some identified principles that should be considered when
designing student team composition in software engineering courses.

\textbf{CCS CONCEPTS}

Applied computing\textasciitilde Collaborative learning •Software and
its engineering

\textbf{Additional Keywords and Phrases:}

Software engineering, education, teamwork, team composition

ACM Reference Format:

First Author's Name, Initials, and Last Name, Second Author's Name,
Initials, and Last Name, and Third Author's Name, Initials, and Last
Name. 2018. The Title of the Paper: ACM Conference Proceedings
Manuscript Submission Template: This is the subtitle of the paper, this
document both explains and embodies the submission format for authors
using Word. In Woodstock '18: ACM Symposium on Neural Gaze Detection,
June 03--05, 2018, Woodstock, NY. ACM, New York, NY, USA, 10 pages.
NOTE: This block will be automatically generated when manuscripts are
processed after acceptance.

\hypertarget{introduction}{%
\section{Introduction}\label{introduction}}

Software is developed by skilled professionals who work in teams.
Software engineering is a collaborative discipline and requires
professionals to learn how to interact with other coworkers who may not
be from the same background. The joint work requires the application of
technical skills and professional practices. Therefore, the software
engineering team members must possess the necessary skills and
competencies.

Teamwork in Software Engineering (SE) education allows students to
enhance their knowledge while working with peers. Students need to learn
the skill to succeed in a competitive university environment. They can
learn and practice teamwork through group assignments, where they should
work and collaborate with their peers. The arrangement is an effective
educational means and helps students learn the phenomenon.

Teamwork should be an integral part of SE coursework as it allows
students to learn and practice skills needed in the software industry
{[}\protect\hyperlink{bib2}{2},\protect\hyperlink{bib3}{3}{]}. They need
to cooperate in the course-related shared tasks for at least two
reasons, first) to solve the given problem, second) to emulate the
professional community that, in turn, helps develop abilities,
encourages motivation, and helps achieve common goals. Therefore,
teamwork is essential for SE students to build professional capabilities
for their future endeavors. That, in turn, prepares them to collaborate
with their future colleagues in the software industry. The application
and consolidation of the knowledge they gained at the university become
accessible in an industrial setting if students have effectively
practiced that during their studies. The phenomenon prepares work-ready
graduates by making them work on SE-related course assignments and
exercises {[}\protect\hyperlink{bib4}{4},\protect\hyperlink{bib5}{5}{]}.

In general, the reason for teamwork is to help achieve tasks that are
beyond the capacity of individuals. Combining individuals into teams can
still not meet the benefits of teamwork if team members are not combined
by considering specific attributes. The reason is that teams require
multiple considerations to operate effectively
{[}\protect\hyperlink{bib1}{1}{]}. Therefore, the assignment of students
to teams to carry out course-related exercises should be carefully done
by SE teachers.

\hypertarget{research-methodology}{%
\section{Research Methodology}\label{research-methodology}}

Action research was chosen as the overall research methodology. At the
beginning of the research project, as a pre-study, a Focus group session
was organized consisting of three experienced SE teachers as experts.
The participating experts had long experience working in the industry
and teaching SE courses to international student groups. In the Focus
group session, student team composition in courses containing group work
assignments emerged as one of the main factors affecting learning
essential teamwork skills.

Team composition in SE course group work assignments was chosen as the
research topic for the ongoing study presented here. Of the ongoing
study, two of the first phases of the Action research cycle, Diagnosing
and Action planning, are presented here. Regarding the objectives of the
Action research study, the research questions for the first two phases,
Diagnosis and Action Planning, were formulated below.

\textbf{RQ-1.} How team composition affects students\textquotesingle{}
teamwork in SE course group assignments?

\textbf{RQ-2.} What team composition attributes should be considered in
SE course group assignment design?

The Action research cycle used is specified based on
{[}\protect\hyperlink{bib6}{6}{]} and adapted for the study as presented
in \protect\hyperlink{fig1}{Figure 1.} In the Diagnosis phase, an
understanding of the aspects affecting teamwork in SE course group
assignments is formed. Based on that, improvement needs and
possibilities are outlined. In the Action Planning phase, essential team
composition attributes are identified and potential suggestions for
action implementation in the later phases of Action research cycle are
outlined.

In the study, the Diagnosing phase was done in two steps involving
different research actions. The first step was empirical, based on
analysis of the earlier mentioned Focus group data obtained from
experts. It deals with familiarizing oneself with the problem and
entails deciding on a problem that is confronted and needs be
addressed\textbf{.~}In the given case, it could be a situation critical
for teaching. Therefore, teamwork in software engineering education was
studied in this step of the Diagnosing phase to identify the practical
development needs and factors affecting those. The second step was done
based on the literature.

\includegraphics[width=3.32708in,height=1.55972in]{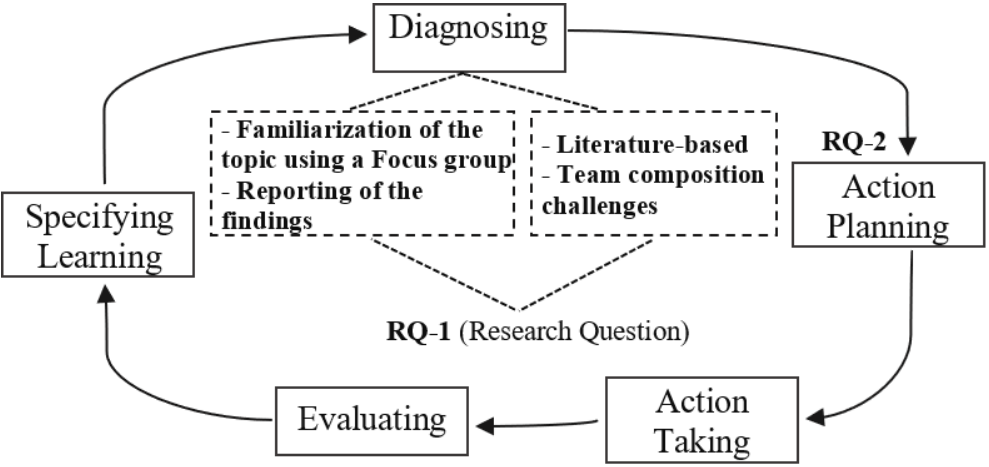}

\protect\hypertarget{fig1}{}{}Figure 1: Action Research Cycle of the
Study

\hypertarget{results}{%
\section{Results}\label{results}}

In order to answer the first research question of the study, the first
step within the diagnosis phase identifies the problem domain into
multiple themes of team composition factors that were identified from
the expert opinion on the subject. Those attributes include homogeneity
or heterogeneity, team size, culture, and personal characteristics such
as age, gender, and ability. There exist multiple types of diversity,
for instance, age, level of skills, gender, culture, and academic
background of the team members.

Student team composition in SE education is situation dependent. There
does not exist any single solution for all cases and conditions.
Multiple solutions exist, depending on the case, and it is more
beneficial to vary the team composition design between consecutive
courses in the curriculum.

The second research question addresses the Action research
cycle\textquotesingle s Action Planning phase: what team composition
attributes should be considered in the group assignment design? As a
result of the study, some relevant planning principles were discovered
to be significant for team composition design. Those principles will be
applied in the next phase of the Action research study. The identified
principles for team composition are the following.

\begin{enumerate}
\def\labelenumi{\arabic{enumi}.}
\item
  The student team composition should be designed by the teachers rather
  than allowing the students to form the teams by themselves. In team
  composition design, various diversity attributes, which are relevant
  in the context, should be considered. Homogenous teams should be
  avoided When the purpose is that the students would learn to work in
  different kinds of teams in their future professional careers.
\item
  In international education, the teams should consist of students with
  internationally diverse backgrounds. When teams consist of students
  from different cultures, they are learning, in addition to the
  substance of the course, while also understanding and capabilities for
  working in international teams. That is valuable in the present-day SE
  working environment, which is highly international.
\item
  Gender balance should also be taken into consideration in team
  composition. In general, teams including representation from each
  gender are preferred. However, it can also cause some issues in
  certain situations, especially with young student groups.
\item
  If the assigned team is homogenous with respect to skills and
  educational background, the group can work more effectively and
  productively with fewer conflicts and dissatisfaction. However, the
  variation in learning between the teams is higher. In addition, the
  students will be missing the learning capabilities to work in
  heterogeneous teams, which is the real-world requirement for their
  professional career.
\end{enumerate}

When student team composition is designed for the course assignment by
the teacher, relevant information about the utilized team composition
attributes needs to be available. This lack of that information can be a
limiting factor in the team composition design. In practice, the teacher
needs to rely only on available information.

\hypertarget{conclusion}{%
\section{Conclusion}\label{conclusion}}

The ongoing study presented in this paper investigates the challenges
related to team composition in software engineering education. According
to the study, team composition can influence students\textquotesingle{}
teamwork in multiple ways. For instance, diversity in its many forms can
make teams underperform through a lack of understanding and
communication among peers and lead to feelings of conflict or
dissatisfaction. The current findings from the investigation of the
research questions suggest that multiple team composition attributes
concerning the heterogeneity of the team should be considered in the SE
group assignment design. Diversity attributes such as gender, culture,
skill level, and academic background of individual team members can
affect team operation, productivity, and learning during teamwork. The
teachers should also consider the need for situation dependence and
variation in the composition of student teams for different courses.

REFERENCES

\protect\hypertarget{bib1}{}{}\textless bib
id="bib1"\textgreater\textless number\textgreater{[}1{]}\textless/number\textgreater Daniel
Levi and David A. Askay. 2020. \emph{Group dynamics for teams}. Sage
Publications.\textless/bib\textgreater{}

\protect\hypertarget{bib2}{}{}\textless bib
id="bib2"\textgreater\textless number\textgreater{[}2{]}\textless/number\textgreater Kurt
Schneider, Olga Liskin, Hilko Paulsen, and Simone Kauffeld. 2015. Media,
mood, and meetings: Related to project success? \emph{ACM Transactions
on Computing Education (TOCE)} 15, no. 4:
1-33.\textless/bib\textgreater{}

\protect\hypertarget{bib3}{}{}\textless bib
id="bib3"\textgreater\textless number\textgreater{[}3{]}\textless/number\textgreater Mary
Shaw, Jim Herbsleb, and Ipek Ozkaya. 2005. Deciding what to design:
Closing a gap in software engineering education. In \emph{Proceedings of
the 27th international Conference on Software Engineering}, pp.
607-608.\textless/bib\textgreater{}

\protect\hypertarget{bib4}{}{}\textless bib
id="bib4"\textgreater\textless number\textgreater{[}4{]}\textless/number\textgreater Arabella
Volkov and Michael Volkov. 2015. Teamwork benefits in tertiary
education: Student perceptions that lead to best practice assessment
design. \emph{Education and Training} 57(3).\textless/bib\textgreater{}

\protect\hypertarget{bib5}{}{}\textless bib
id="bib5"\textgreater\textless number\textgreater{[}5{]}\textless/number\textgreater Bernd
Bruegge, Stephan Krusche, and Lukas Alperowitz. 2015. Software Engi-
neering Project Courses with Industrial Clients. ACM TOCE 15, 4 (2015),
17.\textless/bib\textgreater{}

\protect\hypertarget{bib6}{}{}\textless bib
id="bib6"\textgreater\textless number\textgreater{[}6{]}\textless/number\textgreater Gerald
I. Susman and Roger D. Evered. 1978. An Assessment of the Scientific
Merits of Action Research. \emph{Administrative Science Quarterly}, vol.
23, no. 4, 1978, pp. 582--603. \emph{JSTOR},
\url{https://doi.org/10.2307/2392581}\textless/bib\textgreater{}

\end{document}